\documentclass[notitlepage]{revtex4-1}
\pdfoutput=1
\usepackage{amsmath}
\usepackage{eurosym}
\usepackage{graphicx}
\usepackage{color, soul}
\usepackage{notoccite}
\usepackage{mathtools}
\usepackage{bm}

\begin{document}

\title{Statistical Mechanics and Hydrodynamics of Self-Propelled Hard Spheres%
}
\author{Benjamin Hancock}
\email{bhancock@brandeis.edu}
\author{Aparna Baskaran}
\email{aparna@brandeis.edu}
\date{\today }

\begin{abstract}
Starting from a microscopic model of self-propelled hard spheres we use
tools of non-equilibrium statistical mechanics and the kinetic theory of
hard spheres to derive a Smoluchowski equation for interacting Active
Brownian particles. We illustrate the utility of the statistical mechanics
framework developed with two applications. First, we derive the steady state
pressure of the hard sphere active fluid in terms of the microscopic
parameters and second, we identify the critical density for the onset of motility-induced phase separation in this system. We show that both these quantities agree well with overdamped simulations of active Brownian particles with excluded volume interactions given by steeply repulsive potentials. The results presented here can be used to incorporate excluded volume effects in diverse models of self-propelled particles.
\end{abstract}

\maketitle


\affiliation{ Martin Fisher School of Physics, Brandeis University, Waltham, MA 02453, USA
}



\section{Introduction}

In recent years, the study of active materials has been on the forefront of
research in soft condensed matter physics. A particular class of active
materials is one composed of self propelled particles that convert energy
from a local bath into persistent motion. Self propelled particles describe
systems on many length scales, ranging from bacteria \cite{Cates2012a, berg}%
, synthetic colloids \cite{ginot, Hong2007, Palacci2013}, vibrated granular
rods \cite{menon}, to schools of fish \cite{Marchetti2013, Ballerini29012008}%
. Despite the fact that there exists continual energy consumption and
dissipation at the level of individual particles, collections of
self-propelled particles form large-scale, stable, coherent structures \cite%
{Sriram}. Phenomena exhibited include athermal phase separation among purely
repulsive particles \cite{Buttinoni2013, Cates2014a, Redner2013, Eli2014},
anomalous mechanical properties \cite{Yang2014, takatori, Solon2014, mallory}%
, emergent structures and pattern formation \cite{hagan, gopinath, farrell}.

Theoretical progress in understanding active materials has been propelled by
the study of minimal models for self-propelled particles. The most widely
studied of these minimal models is known as an active Brownian particle
(ABP). ABPs travel with constant velocity along their body axis and the
orientation of their body axis changes through ordinary rotational
diffusion. While numerical investigations of ABPs abound in the literature
and have led to a significant understanding of the dynamics of this system,
analytical progress for interacting particles has been difficult. The reason
that analytical descriptions of the collective behavior remain elusive is
that interactions have a finite collision time, which in turn would result
in intractable many body effects. To overcome this challenge, authors have
taken phenomenological approaches such as incorporating the effect of
interactions in a density dependence of the self propulsion speed \cite%
{Cates2014a, Cates2013, Cates2012a, Tailleur2009, Solon2015rtp} and a mean
field version of dynamical density functional theory \cite%
{bialke,bialke2,bialke3}.

In this paper, we aim to capture these many body effects by applying the
well developed theoretical framework of hard-sphere liquids \cite{ernst,
hansen} to a fluid consisting of ABPs. Starting with underdamped Langevin
equations to describe the dynamics of self-propelled particles that interact
with hard core instantaneous collisions, we systematically coarse grain the
microdynamics to obtain a description applicable on longer length and time
scales after which we take the limit of large friction to obtain an
effective description for an overdamped systems of ABPs. The primary
theoretical result in this work is a first principles derivation of the
statistical mechanics of self-propelled hard spheres. To illustrate the
utility of this result, we compute two well studied emergent properties in
fluids of ABPs: the mechanical pressure and the phase boundary that
determines the onset of athermal phase separation into a dense liquid and a
dilute gas. We then compare our results to those in the literature obtained
from numerical studies and more phenomenological theoretical approaches and
use this comparison to place the context of our work within the existing
body of work.

\section{Theoretical Framework}

\subsection{Microdynamics}

We consider self propelled hard disks in two dimensions with particle
diameter $\sigma $, unit mass, and moment of inertia $I$. The microstate of $%
N$ such hard disks is given by $\Gamma =\{\mathbf{r}_{1}(t),\dots ,\mathbf{r}%
_{n}(t),\mathbf{v}_{1}(t),\dots ,\mathbf{v}_{n}(t)
,\omega _{1}(t),\dots ,\omega _{n}(t),\theta _{1}(t),\dots ,\theta _{n}(t)\}$%
. Here the variables ($\mathbf{r}_{i}$, $\mathbf{v}_{i}$, $\theta _{i}$, $%
\omega _{i}$) are the position, velocity, orientation, and angular velocity
respectively of the $i$ th particle. The equations of motion in this case
are given by $\partial _{t}\mathbf{r}_{i}=\mathbf{v}_{i}$ and $\partial
_{t}\theta _{i}=\omega _{i}$ where the linear and angular velocities evolve
according to
\begin{equation}
\frac{\partial \mathbf{v}_{i}}{\partial t}=\sum_{j\neq i}T(i,j)\mathbf{v}%
_{i}+F\hat{\mathbf{u}}_{i}-\nabla V(\mathbf{r})-\xi \mathbf{v}_{i}+%
\boldsymbol{\eta }_{i}(t)  \label{1}
\end{equation}%
\begin{equation}
\frac{\partial \omega _{i}}{\partial t}=-\xi _{R}\omega _{i}+\eta _{i}^{R}(t).
\label{2}
\end{equation}%
Here, the unit vector $\hat{\mathbf{u}}=(\cos (\theta ),\sin (\theta ))$ is
the orientation of the particle's body axis along which a propulsive force
of strength $F$ acts and $V(\mathbf{r})$ is some external potential. The
binary collision operator $T(i,j)$ generates the instantaneous linear
momentum transfer between disks at contact and is given by \cite{ernst}
\begin{equation}
T(i,j)=\sigma \int d\hat{\boldsymbol{\sigma }}\Theta (-\mathbf{V}_{ij}\cdot
\hat{\boldsymbol{\sigma }})|\mathbf{V}_{ij}\cdot \hat{\boldsymbol{\sigma }}%
|\delta (\mathbf{r}_{ij}-\boldsymbol{\sigma })(b_{ij}-1)  \label{3}
\end{equation}%
where $\hat{\boldsymbol{\sigma }}$ is the unit normal at the point of
contact directed from disk $j$ to disk $i$ and $\mathbf{V}_{ij}=\mathbf{v}%
_{i}-\mathbf{v}_{j}$. The operator $b_{ij}$ replaces pre-collisional
velocities with post-collisional velocities, e.g., $b_{12}\mathbf{v}_{1}=%
\mathbf{v}_{1}-(\mathbf{V}_{12}\cdot \hat{\boldsymbol{\sigma }})\hat{%
\boldsymbol{\sigma }}$ $\ $for collisions which conserve energy and
momentum. The spatial delta function ensures that particles are in contact.
The prefactors that depend on the relative velocities $\mathbf{V}_{ij}$
ensures that the incoming flux of colliding particles is taken into account
correctly. The random forces $\boldsymbol{\eta }_{i}(t)$ and $\eta
_{i}^{R}(t)$ are Gaussian white noise variables with correlations given by $%
\langle \eta _{i}^{\alpha }(t)\eta _{j}^{\beta }(t^{\prime })\rangle
=2k_{B}T\xi \delta _{\alpha \beta }\delta _{ij}\delta (t-t^{\prime })$ and $%
\langle \eta _{i}^{R}(t)\eta _{j}^{R}(t^{\prime })\rangle =2k_{B}T_{R}\xi
_{R}\delta _{ij}\delta (t-t^{\prime })$ respectively. Latin indices label
the particle number, while Greek indices label the vector components of the
noise. The noise amplitudes depend on parameters $T,T_{R}$ that have the
units of temperature. These parameters need not be the same as $T_{R}$ is an
intrinsic quantity describing the reorientation of the active drive.

Thus, Eqs.(\ref{1}-\ref{2}) are the Langevin equations for $N$ interacting self-propelled
hard disks. For a single particle in the high friction limit these equations
would reduce to the well studied \cite%
{Redner2013,Fily2012,Cates2012a,Romanczuk2012, Cates2014a, Marchetti2013,
Yang2014, takatori, Solon2014, mallory, Farage, maggi, Cates2013,
Solon2015rtp, bialke, Tailleur2009} overdamped Langevin equations $\partial
_{t}\mathbf{r}=v_{0}\hat{\mathbf{u}}+\bar{\boldsymbol{\eta }}$ and $\partial
_{t}\theta =\bar{\eta}^{R}$ where the self-propulsion velocity is given by $%
v_{0}=F/\xi $ and the noise terms are given by $\bar{\boldsymbol{\eta }}=%
\boldsymbol{\eta }/\xi $ and $\bar{\eta}^{R}=\eta ^{R}/\xi _{R}$.

\subsection{Statistical Mechanics}

\noindent \underline{\textbf{Illustration of Coarse-graining Technique}}: We
seek to derive the statistical mechanics of this system of self-propelled
hard disks in the large friction limit. One could derive the statistical
mechanics from the overdamped Langevin equation mentioned at the end of the
previous subsection rather straightforwardly. In this work, we are using
hard disks to be able to tractably incorporate the physics of excluded
volume interactions into the statistical mechanics. In this case, the large
friction limit needs to be taken at the level of the Fokker-Planck equation.
In order to illustrate the technique involved, let us begin by considering a
collection of noninteracting particles (described by Eqs.(\ref{1}-\ref{2}) without the
binary collision operator). In this case the statistical mechanics is given
by the one particle probability distribution function $f(\mathbf{r},\theta ,%
\mathbf{v},\omega ,t)$ of finding a particle with some position $\mathbf{r}$%
, orientation $\theta $, velocity $\mathbf{v}$, and angular velocity $\omega
$ at time $t$. This PDF obeys the Fokker-Planck equation given by \cite%
{aparna2010}
\begin{equation}
\partial _{t}f(\mathbf{r},\theta ,\mathbf{v},\omega ,t)+\mathcal{D}f(\mathbf{%
r},\theta ,\mathbf{v},\omega ,t)=0  \label{4}
\end{equation}%
and the Fokker-Planck operator $\mathcal{D}$ is given by
\begin{equation}
\mathcal{D}=\mathbf{v}\cdot \nabla _{\mathbf{r}}+\omega \partial _{\theta }+F%
\hat{\mathbf{u}}\cdot \nabla _{\mathbf{v}}-\nabla _{\mathbf{r}}V(\mathbf{r}%
)\cdot \nabla _{\mathbf{v}}-\xi \nabla _{\mathbf{v}}\cdot \mathbf{v}-\xi
_{R}\partial _{\omega }\omega -k_{B}T\xi \nabla _{\mathbf{v}%
}^{2}-k_{B}T_{R}\xi _{R}\partial _{\omega }^{2}.  \label{5}
\end{equation}%
Let us define the particle concentration $c(\mathbf{r},\theta ,t)=\int d%
\mathbf{v}d\omega f(\mathbf{r},\theta ,\mathbf{v},\omega ,t)$, a
translational current $\mathbf{J}^{T}=\int d\mathbf{v}d\omega \mathbf{v}f(%
\mathbf{r},\theta ,\mathbf{v},\omega ,t)$, and a rotational current $%
J^{R}=\int d\mathbf{v}d\omega \omega f(\mathbf{r},\theta ,\mathbf{v},\omega
,t)$. By taking appropriate velocity moments of Eq.(\ref{4}), one finds that
the concentration field obeys a conservation law of the form
\begin{equation}
\partial _{t}c+\nabla _{\mathbf{r}_{1}}\cdot \mathbf{J}^{T}+\partial
_{\theta }J^{R}=0.  \label{6}
\end{equation}%
In the large friction limit, the currents are given by
\begin{subequations}
\begin{align}
& \lim_{t\gg 1/\xi }\mathbf{J}^{T}=-\frac{1}{\xi }\big[-F\hat{\mathbf{u}}%
c+c\nabla V(\mathbf{r})+\nabla \cdot \langle \mathbf{v}\otimes \mathbf{v}%
\rangle +\partial _{\theta }\langle \omega \mathbf{v}\rangle \big] \label{7a} \\
& \lim_{t\gg 1/\xi _{R}}J^{R}=-\frac{1}{\xi _{R}}\big[\nabla \cdot \langle
\omega \mathbf{v}\rangle +\partial _{\theta }\langle \omega ^{2}\rangle \big] \label{7b}
\end{align}%
where the brackets in the above represent averages over linear and angular
velocities. In order to have a closed equation for the concentration we must
evaluate the averages in Eqs.(\ref{7a}-\ref{7b}). We assume the PDF can be written as
\end{subequations}
\begin{equation}
f(\mathbf{r},\theta ,\mathbf{v},\omega ,t)=\frac{1}{2\pi k_{B}T}\sqrt{\frac{1%
}{2\pi k_{B}T_{R}}}c(\mathbf{r},\theta ,t)e^{-\frac{(\mathbf{v}-v_{0}\hat{%
\mathbf{u}})^{2}}{2k_{B}T}}e^{-\frac{\omega ^{2}}{2k_{B}T_{R}}}
\end{equation}%
with $v_{0}=F/\xi $. This assumption says that in this large friction
regime, there exists separation of time scales for spatial relaxations and
velocity relaxations, the latter being fast and so at late times, the
velocity distribution has relaxed to its local equilibrium form \cite%
{aparna2010}. The velocity averages can then be evaluated and the dynamical
equation reads
\begin{equation}
\partial _{t}c+v_{0}\hat{\mathbf{u}}\cdot \nabla c-\xi ^{-1}\nabla \cdot %
\big[c\nabla V(\mathbf{r})\big]-\nabla \cdot \tensor{\mathbf{D}}\cdot \nabla
c-D_{r}\partial _{\theta }^{2}c=0  \label{9}
\end{equation}%
In the above $D_{r}=\frac{k_{B}T_{R}}{\xi _{r}}$ and the diffusion tensor is
of the following form $D_{\alpha \beta }=D_{\parallel }\hat{u}_{\alpha }\hat{%
u}_{\beta }+D_{\perp }(\delta _{\alpha \beta }-\hat{u}_{\alpha }\hat{u}%
_{\beta })$. With $D_{\parallel }=v_{0}^{2}\xi ^{-1}+k_{B}T\xi ^{-1}$ and $%
D_{\perp }=k_{B}T\xi ^{-1}$. Note that in the completely thermal limit we
recover ordinary isotropic diffusion $D_{\alpha \beta }=\xi
^{-1}k_{B}T\delta _{\alpha \beta }=D_{t}\delta _{\alpha \beta }$. In the
purely self-propelled limit we obtain $D_{\alpha \beta }=\xi ^{-1}v_{0}^{2}%
\hat{u}_{1\alpha }\hat{u}_{1\beta }$ as was reported in \cite{aparna2010}
for self-propelled hard rods. This procedure yields the well studied
Smoluchowski equation for ABPs that others have obtained \cite%
{Romanczuk2012,Solon2015rtp,bialke}.

\noindent \underline{\textbf{Result}}: Now, repeating the calculation but
retaining the hard core interactions (see Appendix A), the Smoluchowski
equation is of the following form
\begin{equation}
\partial _{t}c+\nabla _{\mathbf{r}}\cdot \mathbf{J}^{T}-D_{r}\partial
_{\theta }^{2}c=0  \label{10}
\end{equation}%
Note that the rotational part of the Smoluchowksi equation remains unchanged
because no torques are exerted in a collision of smooth disks. However, the
translational current has the form
\begin{equation}
\mathbf{J}^{T}=v_{0}\hat{\mathbf{u}}c-\xi ^{-1}c\nabla V(\mathbf{r})-%
\tensor{\mathbf{D}}\cdot \nabla c+\xi ^{-1}\big(\mathbf{I}^{thermal}+\mathbf{%
I}^{sp}+\mathbf{I}^{cross}\big).  \label{11}
\end{equation}%
The first three terms of Eq.(\ref{11}) are equivalent to the terms in Eq.(\ref{9})
and the additional terms in Eq.(\ref{11}) are collisional contributions to the
translational current. They are given by
\begin{subequations}
\begin{align}
& I_{\alpha }^{thermal}=\frac{4\pi ^{2}\sigma (k_{B}T)^{3}}{(2\pi
k_{B}T+v_{0}^{2})^{2}}\int_{\hat{\boldsymbol{\sigma }},\theta _{2}}c^{(2)}(%
\mathbf{r}_{1},\theta _{1},\mathbf{r}_{1}-\boldsymbol{\sigma },\theta _{2},t)%
\hat{\sigma}_{\alpha }  \label{12a} \\
& I_{\alpha }^{sp}=\frac{\sigma v_{0}^{6}}{(2\pi k_{B}T+v_{0}^{2})^{2}}\int_{%
\hat{\boldsymbol{\sigma }},\theta _{2}}\Theta (-\hat{\mathbf{u}}_{12}\cdot
\hat{\boldsymbol{\sigma }})(\hat{\mathbf{u}}_{12}\cdot \hat{\boldsymbol{%
\sigma }})^{2}c^{(2)}(\mathbf{r}_{1},\theta _{1},\mathbf{r}_{1}-\boldsymbol{%
\sigma },\theta _{2},t)\hat{\sigma}_{\alpha } \label{12b}\\
& I_{\alpha }^{cross}=\frac{\sigma }{(2\pi k_{B}T+v_{0}^{2})^{2}}\int_{\hat{%
\boldsymbol{\sigma }},\theta _{2}}c^{(2)}(\mathbf{r}_{1},\theta _{1},\mathbf{%
r}_{1}-\boldsymbol{\sigma },\theta _{2},t) \label{12c}
\end{align}%
\end{subequations}
\begin{equation*}
\times \lbrack 2\pi (k_{B}T)^{2}v_{0}^{2}+\pi k_{B}Tv_{0}^{4}((\hat{\mathbf{u%
}}_{1}\cdot \hat{\boldsymbol{\sigma }})^{2}+(\hat{\mathbf{u}}_{2}\cdot \hat{%
\boldsymbol{\sigma }})^{2})-\sqrt{8\pi (k_{B}T)^{3}}v_{0}^{3}\hat{\mathbf{u}}%
_{12}\cdot \hat{\boldsymbol{\sigma }}]\hat{\sigma}_{\alpha }
\end{equation*}%
and where $\hat{\mathbf{u}}_{12}=\hat{\mathbf{u}}_{1}-\hat{\mathbf{u}}_{2}$.

\noindent \underline{\textbf{Discussion}}:

\begin{enumerate}
\item The above collisional contributions in Eq.(\ref{10}) all have the form
of a mean field force, $I_{\alpha }=\int_{\hat{\boldsymbol{\sigma }},\theta
_{2}}F_{\alpha }(\theta _{1},\theta _{2},\hat{\boldsymbol{\sigma }})c^{(2)}(%
\mathbf{r}_{1},\theta _{1},\mathbf{r}_{1}-\boldsymbol{\sigma },\theta
_{2},t) $. Where $F_{\alpha }(\theta _{1},\theta _{2},\hat{\boldsymbol{%
\sigma }})$ is some orientation dependent force density and $c^{(2)}(\mathbf{%
r}_{1},\theta _{1},\mathbf{r}_{1}-\boldsymbol{\sigma },\theta _{2},t)$ is
the two body probability distribution function of finding particle 1 with
position $\mathbf{r}_{1}$ and orientation $\theta _{1}$ and particle 2 with
position $\mathbf{r}_{1}-\boldsymbol{\sigma }$ and orientation $\theta _{2}$
at a time $t$ (see fig.\ref{fig1}).
\begin{figure}[h]
\includegraphics[width=6.5cm,height=6cm]{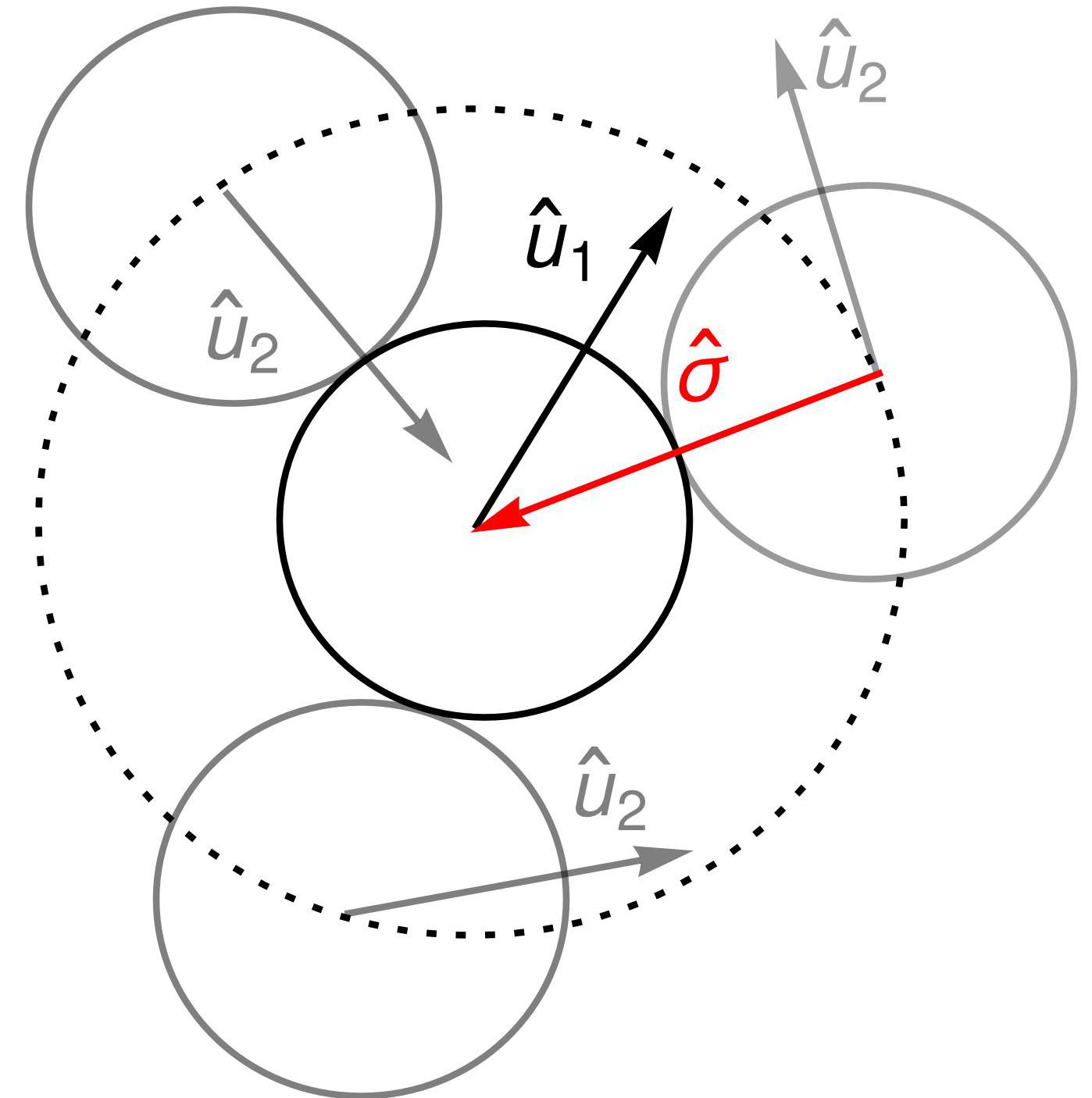}
\caption{(color online) An illustration of a collision between particle 1
with orientation $\hat{\mathbf{u}}_{1}(\protect\theta _{1})$ and particle 2
with orientation $\hat{\mathbf{u}}_{2}(\protect\theta _{2})$ . The unit
normal vector $\hat{\boldsymbol{\protect\sigma }}$ at the point of contact
is directed from particle 2 to particle 1. A collision can occur when the
center of particle 2 lies anywhere on the dotted line, as illustrated by the
gray spheres. The mean field force is obtained by integrating over all
configurations $\hat{\mathbf{u}}_{1}$,$\hat{\mathbf{u}}_{2}$ and $\hat{%
\boldsymbol{\protect\sigma }}$}
\label{fig1}
\end{figure}

\item In the completely thermal limit ($v_{0}=0$) the only term that would
contribute to the collision integral is Eq.(\ref{12a}) and would reproduce
the statistical mechanics of thermal hard spheres in the overdamped limit.
This is precisely the contribution one would find starting from the revised
Enskog theory \cite{balescu1, kinetic,chapman}.

\item Eq.(\ref{12b}) is a contribution from collisions arising from self-propulsion
alone. Within this equation is a theta function which constrains the range
of allowed orientations of the self-propulsion direction of the second
particle. That is, the theta function is nonzero only for orientations that
would result in a collision.

\item Eq.(\ref{12c}) is a collisional contribution that arises from the coupling of
thermal noise and self-propulsion and contains the leading order
contribution to the translational current from the collisions among
particles (see Appendix A).
\end{enumerate}

In this section we have outlined the systematic derivation of the
statistical mechanics of self-propelled hard spheres. The results described
above should be useful to describe any collection of active particles that
interact through strongly repulsive short range interactions, as will be
shown in later sections. We now illustrate the applicability of the derived
statistical mechanics by investigating some macroscopic properties of a
fluid of active particles.

\section{Steady State Mechanical Properties}

\begin{figure}[h]
\includegraphics[width=10cm,height=5.5cm]{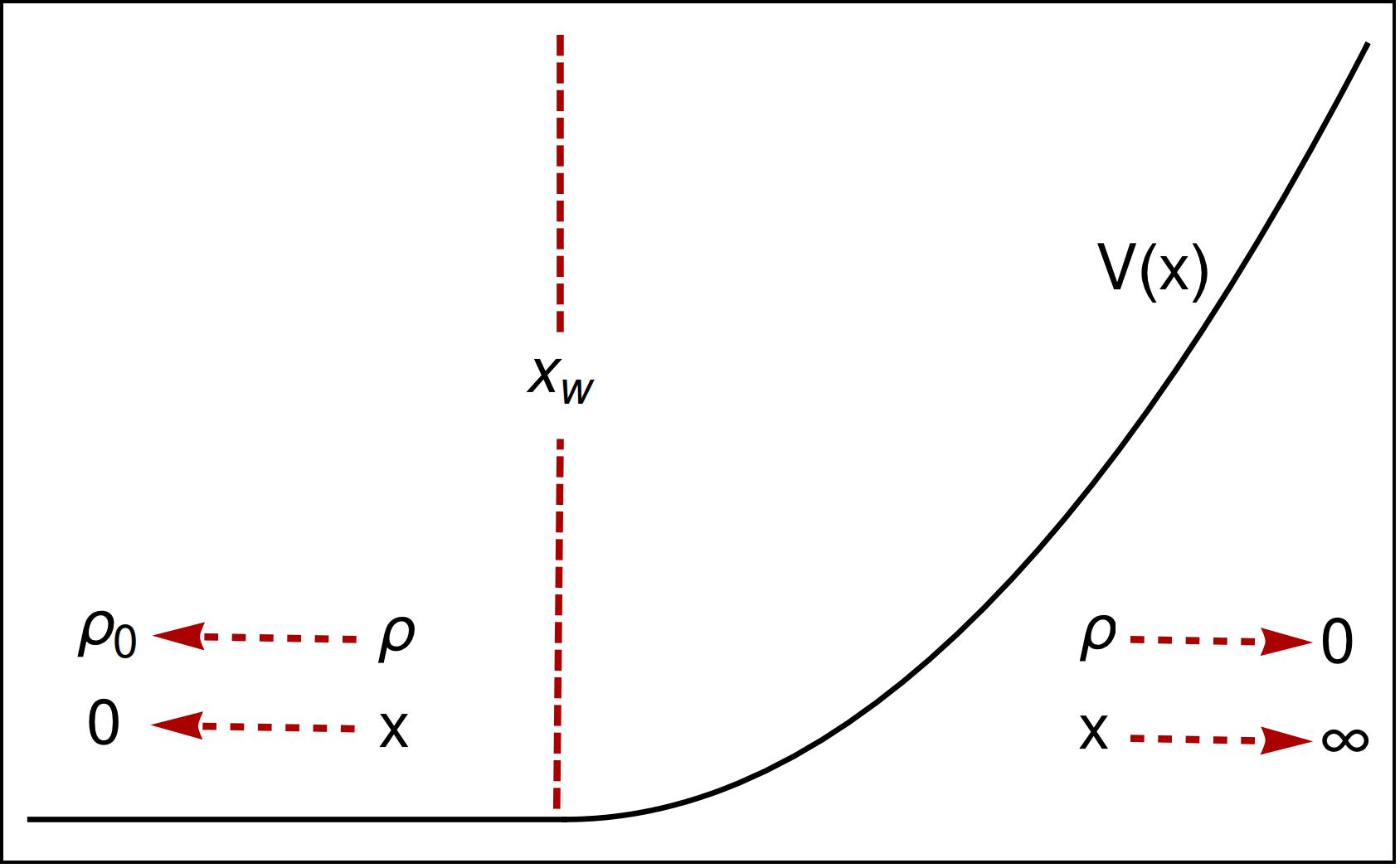}
\caption{(color online) An illustration of the confining potential}
\label{fig2}
\end{figure}

As a first illustration, we seek to use the statistical mechanics developed
above to calculate a steady state property of this system. Let us consider
the pressure of an active fluid. As has been shown in \cite{Solon2014} this
is indeed a state variable for self-propelled particles in the absence of
any torques as is the case for smooth hard disks considered here.
Mechanically one can then define pressure by considering the system as being
confined by a wall at some position $x_{w}\gg 0$ (see fig.\ref{fig2}).
represented by some confining potential $V(x)$ and where $x=0$ is taken to
be deep in the bulk of the active particle fluid. We assume that deep in the
bulk of the fluid the density has a constant bulk value $\rho _{0}$, far
beyond $x_{w}$ the density vanishes, and that the system stays uniform in
the $\hat{y}$-direction. By Newton's 3rd Law, the pressure can be computed
from the total force acting on the wall,
\begin{equation}
P=\int_{0}^{\infty }\rho (x)\partial _{x}V(x)dx.  \label{14}
\end{equation}%
Using the procedure developed in \cite{Solon2014} (see Appendix B for
details), we arrive at the following expression for the pressure
\begin{equation}
P=P_{0}+\int d\theta \bigg[\frac{v_{0}}{D_{r}}\hat{u}%
_{x}(I_{x}^{thermal}+I_{x}^{sp}+I_{x}^{cross})\bigg|_{x=0}+\int_{0}^{\infty
}(I_{x}^{thermal}+I_{x}^{sp}+I_{x}^{cross})dx\bigg].  \label{P1}
\end{equation}%
In the above, $P_{0}=\big(\frac{v_{0}^{2}\xi }{2D_{r}}+\frac{v_0^2}{2}+k_BT%
\big)\rho _{0}$ is the pressure of noninteracting
self-propelled particles and $I$'s are the collision kernels given in Eqs.
(\ref{12a}-\ref{12c}).

To make any further analytic progress we must evaluate integrals over the
collisional contributions which involve the two body distribution $c^{\left(
2\right) }$. We now make the ansatz that this two body distribution can be
written as a functional of the one body distributions in the following way,
\begin{equation}
c^{(2)}(\mathbf{r}_{1},\theta _{1},\mathbf{r}_{2},\theta _{2},t)=g(\mathbf{r}%
_{1},\mathbf{r}_{2}\big|\rho (\mathbf{r},t))c(\mathbf{r}_{1},\theta _{1},t)c(\mathbf{r}%
_{2},\theta _{2},t),  \label{c2}
\end{equation}%
where  $g(\mathbf{r}_{1},\mathbf{r}_{2}\big|\rho (\mathbf{r},t))$  is a functional of
the density field. For thermal hard spheres, this function $g$ is the
equilibrium pair correlation function and it provides an exact functional
relationship between the one and two body distributions. In the case of
self-propelled particles, this ansatz cannot be exact as we expect
orientational correlations to play some role. Such orientational
correlations were in fact characterized in \cite{bialke} through the use of
density functional theory and simulations. A systematic estimation of $g$,
even in the limited form we have chosen is a hard problem that is beyond the
scope of the present work. In the following, we use the form of $g$
associated with thermal hard spheres at contact. That is, we assume
orientational correlations can be neglected and that positional correlations
are accounted for in the same way as for thermal hard spheres. The test of
the validity of this assumption will be the comparison to numerical
simulations considered later in this presentation. For the rest of the paper
we use the well known estimate of the contact pair-correlation function
known as the Carnahan-Starling pair-correlation function in 2 dimensions
\cite{balescu1}.
\begin{equation}
g\left( \mathbf{r}_{1},\mathbf{r}_{2}\big|\rho (\mathbf{r},t)\right) =g(\sigma |\rho )=%
\frac{1-\frac{7}{16}\phi }{(1-\phi )^{2}},
\end{equation}%
where $\phi =\frac{\pi }{4}\rho \sigma ^{2}$ is the packing fraction.
This estimate is known to give an accurate description of the fluid phase of hard spheres and consequently, this approximation will not capture crystallization effects. We also note that one can choose other estimates of the pair correlation function, such as the Hypernetted Chain \cite{balescu1}, to approximate density correlations in different parameter regimes. With the ansatz above, the computation of the pressure in Eq.(\ref{P1}) reduces to integrals over the one particle distribution function $c\left(
r,\theta \right) $ which is in turn the steady state solution to the
Smoluchowski equation Eq.(\ref{10}). Using the standard procedure \cite%
{Solon2015rtp, aparnarod} of representing the distribution as a harmonic
expansion in terms of the angular moments, $c=\frac{\rho }{2\pi }+\frac{1}{%
\pi }\mathbf{P}\cdot \hat{\mathbf{u}}+ \tensor{\mathbf{Q}}:(\hat{\mathbf{u}}\hat{\mathbf{u}}-\frac{1}{2}\mathbf{I})+\dots $ where $\mathbf{P}=\int d\theta \hat{\mathbf{u}}c$%
, is the first moment, $\tensor{\mathbf{Q}}=\int d\theta(\hat{\mathbf{u}}\hat{\mathbf{u}}-\frac{1}{2}\mathbf{I})c$ is the second moment, and assuming a
low-moment closure (i.e., truncating the moment expansion at some order, see
appendix C for details) we can now evaluate the integrals in Eq.(\ref{P1})
with the result
\begin{equation}
P=P_0+\lambda _{D}g(\sigma )\rho _{0}^{2}-g(\sigma )\frac{\xi v_{0}}{D_{r}}%
\lambda _{I}\rho _{0}^{2}.  \label{pres}
\end{equation}%
In the above, the constants $\lambda _{D}$ and $\lambda _{I}$ control the
strength of the collisional contributions to the pressure and depend on the
microscopic parameters (see appendix D for the explicit forms). In order to
understand the structure of this result, it is useful to consider some
limits of Eq.(\ref{pres}). First, in the absence of the self-propulsion (i.e. $%
v_{0}=0$) we have.
\begin{equation}
P=k_{B}T\rho _{0}\bigg(1+\frac{\pi \sigma ^{2}g(\sigma )}{2}\rho _{0}\bigg)
\end{equation}%
This is precisely what one would find when calculating the pressure for a
hard sphere gas when using Revised-Enskog theory \cite{ap1, garzo}. It
consists of the ideal gas term plus an additional correction due to the hard
core interactions. In the athermal limit (i.e. $k_{B}T=0$) we obtain
\begin{equation}
P=\frac{v_{0}^{2}}{2D_{r}\xi ^{-1}}\rho _{0}+\frac{v_{0}^{2}}{2}\rho _{0}+%
\frac{3\pi }{16}v_{o}^{2}\sigma ^{2}g(\sigma )\rho _{0}^{2}-\frac{2}{3}\frac{%
v_{0}^{3}\sigma }{D_{r}}g(\sigma )\rho _{0}^{2} \label{19}
\end{equation}%
This limit ($k_{B}T=0$) has been simulated in \cite{Solon2014} using the
Weeks-Chandler-Andersen (WCA) potential. Even though this interaction has a
finite collision time we find that our expression captures the computed
pressure well for low to moderate densities (see fig.3), thus illustrating
the validity of our results for systems with short-range strongly repulsive
interactions.
\begin{figure}[h]
\includegraphics[width=17cm,height=5.8cm]{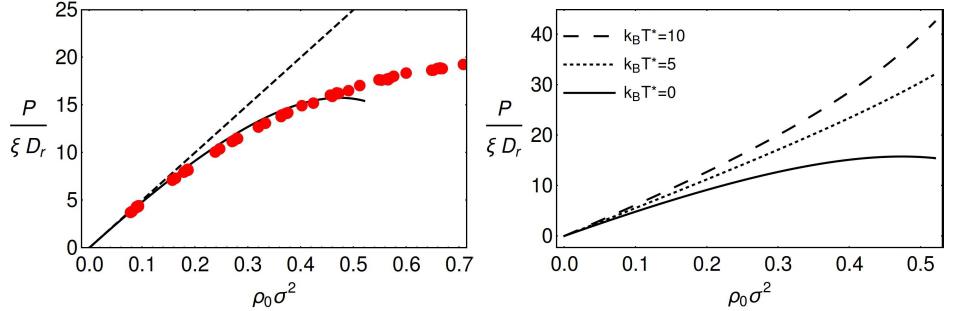}
\caption{(color online) In this figure Eq.(\ref{19}) was made dimensionless by $P=P^{\ast }\protect%
\xi D_{r}$, $v_{0}=v_{0}^{\ast }\protect\sigma D_{r}$, $\protect\rho _{0}=%
\protect\rho _{0}^{\ast }\protect\sigma ^{-2}$, $k_{B}T=k_{B}T^{\ast }%
\protect\xi \protect\sigma ^{2}D_{r}$. Left: The pressure in the athermal
limit. The dots are the simulated results of ABPs interacting via the WCA
potential. The solid black line is the theoretical expression (Eq.(\ref{19})) with a
fit parameter $\protect\alpha =D_{r}/\protect\xi =.027$. The dashed line is
the noninteracting pressure. Right: The pressure when temperature is
included (Eq.(17)) leading to an increase in the pressure as the temperature
is increased.}
\label{fig.3}
\end{figure}

Finally, we compare our result for the pressure with those already in the
literature for an overdamped system of ABPs interacting with repulsive
potentials.  In \cite{Solon2014b}, the authors use fluctuating hydrodynamics
to derive expressions for the pressure of interacting ABPs in terms of
correlation functions between moments of $c(\mathbf{r},\theta ,t)$. They
find that the pressure can be written as the sum of three terms $%
P=P_{0}+P_{I}+P_{D}$ where $P_{0}$ is the ideal active pressure and the
"indirect pressure" $P_{I}$ and "direct pressure" $P_{D}$ are given
respectively by $P_{I}=\frac{v_{0}}{D_{r}}\int d^{2}r^{\prime }F_{x}(\mathbf{%
r}^{\prime })\langle \rho (\mathbf{r}^{\prime })P_{x}(0)\rangle $ and $%
P_{D}=\int_{0}^{\infty }dx\int d^{2}r^{\prime }F_{x}(\mathbf{r^{\prime }}-%
\mathbf{r})\langle \rho (\mathbf{r}^{\prime })\rho (\mathbf{r})\rangle $
where $F_{x}$ is the component of the interaction force between particles
and the angular brackets here represent an average over the noise. In our
theory, we start with a noise averaged dynamical equation. The analogous
contributions for the pressure in our case are as follows. The second term
in Eq. (\ref{P1}) is the indirect pressure $P_{I}$ and is indeed the
evaluation of the corresponding correlation function over the solution of
the Smoluchowski equation for the case of hard spheres.  The third term in
Eq. (\ref{P1}) is the direct pressure $P_{D}$ evaluated for our hard core
interactions. We also note that in \cite{Solon2014b} the authors prove that
the sum $P_{0}+P_{I}$ is equivalent to the "swim pressure" investigated by
\cite{Yang2014,takatori} and the same equivalence holds between our theory
and the swim pressure as well.

Summarizing, in this section, we have used the statistical mechanics of self
propelled hard spheres to tractably evaluate the interaction contributions
to the pressure of an active fluid. We have found good agreement with
numerical simulations of the pressure of ABPs interacting via a steeply
repulsive potential with a finite collision time, thus illustrating the
usefulness of the derived statistical mechanics for a variety of model
active fluids.

\section{Hydrodynamics and Phase Separation}

As a second illustration of the utility of the nonequilibrium statistical
mechanics constructed here, we derive and characterize a dynamical
description of an active fluid on length scales long compared to the
particle diameter and time scale long compared to the mean free time. In
this regime, the relevant dynamical variables are the conserved quantities and quantities associated with any
possible broken symmetries that couple to them. The only conserved quantity is the density of
particles given by the zeroth moment of the probability distribution
\begin{equation}
\rho (\mathbf{r},t)=\int d\theta c(\mathbf{r},\theta ,t).
\end{equation}%
The other relevant dynamical quantities are higher angular moments of concentration field. Of these moments, the only relevant quantity that couples to the hydrodynamic field is the polarization described by the
first moment of the probability distribution.
\begin{equation}
P_{\alpha }(\mathbf{r},t)=\int d\theta \hat{u}_{\alpha }c(\mathbf{r},\theta
,t).
\end{equation}%
We seek to identify the dynamical equations obeyed by these two quantities.

To derive the continuum equations for the relevant macroscopic fields we must take
the corresponding moments of the Smoluchowski equation and they are of the
form
\begin{equation}
\partial _{t}\rho =-\nabla \cdot \mathbf{J}
\end{equation}%
\begin{equation}
\partial _{t}P_{\beta }=-\partial _{\alpha }J_{\alpha \beta }-D_{r}P_{\beta }
\end{equation}%
where
\begin{equation}
\begin{pmatrix}
J_{\alpha } \\
J_{\alpha \beta }%
\end{pmatrix}%
=\int d\theta
\begin{pmatrix}
1 \\
\hat{u}_{\beta }%
\end{pmatrix}%
J_{\alpha }^{T} . \label{Flux}
\end{equation}%
The fluxes in Eq.(\ref{Flux}) are again moments of the two particle
distribution as in the case of the pressure calculation above. In order to
evaluate these fluxes, we proceed as in the preceding section by assuming the
simplest phenomenological closure of the Smoluchowksi equation, that the two
particle distribution function can be written as the product of one particle
distributions $c^{(2)}(\mathbf{r}_{1},\theta _{1},\mathbf{r}_{1}-\boldsymbol{%
\sigma },\theta _{2},t)=g(\sigma )c^{(1)}(\mathbf{r}_{1},\theta
_{1},t)c^{(1)}(\mathbf{r}_{1}-\boldsymbol{\sigma },\theta _{2},t)$ and give
the one particle distribution function a series representation in terms of
its angular moments (Appendix B). As before, we are using the Carnahan-Starling estimate  for $g(\sigma)$. Since we are interested in a long
wavelength description of the system the nonlocal dependence of the
concentration field is expanded in gradients
\begin{equation}
c(\mathbf{r}-\boldsymbol{\sigma },\theta _{2},t)=c(\mathbf{r},\theta
_{2},t)-\sigma _{\alpha }\partial _{\alpha }c(\mathbf{r},\theta
_{2},t)+\dots
\end{equation}%
Using the above gradient expansion coupled with a low moment closure we
arrive the following equations
\begin{equation}
\partial _{t}\rho +v_{0}\nabla \cdot \mathbf{P}-\nabla \cdot \lbrack \mathcal{D}(\rho )\nabla \rho ]+D_{p}\nabla \nabla :\mathbf{P}\mathbf{P}=0
\label{rho}
\end{equation}%
\begin{equation}
\partial _{t}P_{\gamma }+D_{r}P_{\gamma }+\nabla _{\gamma }\mathcal{P}(\rho
)-D_{\perp }\nabla ^{2}P_{\gamma }+2\lambda _{I}\big[P_{\gamma }(\nabla
\cdot \mathbf{P})+(\mathbf{P}\cdot \nabla )P_{\gamma }\big]  \label{P}
\end{equation}
\begin{eqnarray*}
&=&\nabla \cdot \big[\tensor{\mathbf{L}}(P_{\gamma })\nabla \rho \big]%
+\lambda _{2}\big[(\nabla \cdot \mathbf{P})\nabla _{\gamma }\rho +(\mathbf{P}%
\cdot \nabla )\nabla _{\gamma }\rho +\nabla _{\gamma }\left( \mathbf{P}\cdot
\nabla \rho \right) \big] \\
&&-\nabla \cdot \big[K(\rho )\nabla _{\gamma }\mathbf{P}\big]-\nabla
_{\gamma }\big[K(\rho )\nabla \cdot \mathbf{P}\big]
\end{eqnarray*}%
where the explicit expressions for all the macroscopic parameters and the
functions $\mathcal{D}(\rho )$, $K\left( \rho \right) $ and $L_{\alpha \beta
}\left( P_{\gamma }\right) $ in terms of the microscopic parameters of the
model are given in Appendix D.

These macroscopic equations are complex and nonlinear, with the effect of
the repulsive interactions showing up in the coefficients and in the
detailed form of the nonlinearities. While careful study of the phase
behavior predicted by these equations is warranted, we defer this to future
work and make only a few remarks about the structure of these equations. The
density equation above has a similar form to those written down for
non-interacting ABPs but with a density dependent diffusion
coefficient $\mathcal{D}(\rho )$ and a term analogous to the curvature induced
flux term found in hydrodynamic theories of orientable active particles \cite%
{simha}, signifying the fact that the orientation comes with a physical
velocity and hence its fluctuations can result in a diffusive flux. The
polarization equation does not have a homogeneous nonlinearity reflecting
the fact that the interactions among smooth particles are non-aligning, but
still has the complex nonlinearities one expects in Toner-Tu type
hydrodynamic theories \cite{tonertu}. Note that $v_{0}\mathbf{P}$ is a measure of the
collective self-propulsion velocity of the system and thus encompasses a
compressible flow. This is reflected by the presence of the hydrostatic
pressure through $\mathcal{P}=\frac{D_{r}}{v_{0}\xi }P$, together with
additional Euler order terms  $K(\rho )\nabla \cdot \mathbf{P}-\lambda _{2}%
\mathbf{P}\cdot \nabla \rho $ in the dynamical equation for this flow. 

\begin{figure}[h]
\includegraphics[width=7.5cm,height=7.5cm]{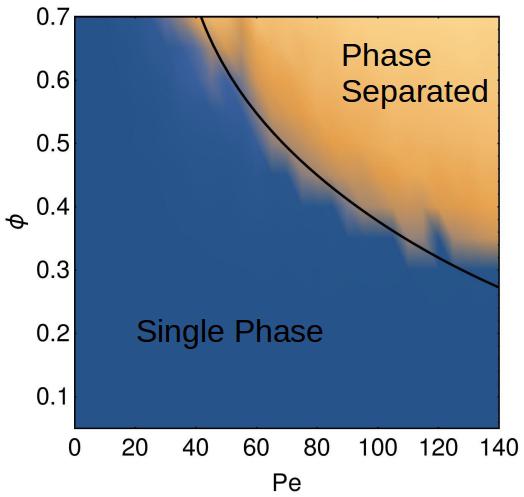}
\caption{(color online) Numerical results for the phase boundary taken from
\protect\cite{Redner2013}. Black line is the critical density at which Eq.(\ref{31})
becomes zero. Here we find good agreement when using $\protect\alpha =.01$
and the dimensionless temperature $k_{B}T^{\ast }=25$. Below the line is the
$\mathcal{D}(\protect\rho )>0$ region, above the line is the $\mathcal{D}(%
\protect\rho )<0$ region.}
\label{fig.4}
\end{figure}

Note that the relaxation of time of Eq.(\ref{P}) is given by $t=1/D_{R}$. In the rest of this section, we focus on the behavior of this system on
times much longer than this characteristic relaxation time. For such times
it is reasonable to assume that the polarization has relaxed to its steady
state value, i.e.,  $\partial _{t}\mathbf{P}=0$. Solving for $\mathbf{P}$ in
Eq.(\ref{P}) and substituting into Eq.(\ref{rho}) we find, neglecting
higher order gradient terms, the following diffusion equation for the
density,
\begin{equation}
\partial _{t}\rho =\nabla \cdot \lbrack D_{eff}(\rho )]\nabla \rho ],
\end{equation}%
where the effective density dependent diffusion coefficient is given by
\begin{equation}
D_{eff}(\rho )=\frac{v_{0}^{2}}{2D_{r}}+\mathcal{D}(\rho )-\frac{2v_{0}}{%
D_{r}}\lambda _{I}\rho .  \label{31}
\end{equation}%
In the limit $k_{B}T\rightarrow 0$,  this effective diffusion coefficient
becomes (putting in the explicit forms for $\mathcal{D}(\rho)$ and $\lambda _{I}$ from
Appendix D)
\begin{equation}
D_{eff}(\rho )=\frac{v_{0}^{2}}{2D_{r}}+\frac{v_{0}^{2}}{2\xi }+g(\sigma
)\frac{3\pi }{8\xi }v_{0}^{2}\sigma ^{2}\rho -g(\sigma )\frac{4}{3}\frac{%
v_{0}^{3}\sigma \rho }{D_{r}\xi }  \label{32}
\end{equation}%
The last term in Eqs.(\ref{31}-\ref{32}) is negative and there exists a
critical density $\rho _{c}$ or equivalently a critical packing fraction $%
\phi _{c}=\rho _{c}\frac{\pi }{4}\sigma ^{2}$ above which this diffusion
coefficient becomes negative. This signals the onset of clustering in the
system that has been referred to in the literature as Motility Induced Phase
Separation (MIPS). Using the Carnahan-Starling estimate (Eq.(24)) for $%
g(\sigma )$ one can identify this critical density as a function of system
parameters and this is shown by the black line in fig.\ref{fig.4}.

To better understand this critical density, it is useful to take some
limits. First consider the limit of high Peclet number, in this limit the
critical density predicted by Eq.(\ref{31}) is of the form to $\phi _{c}=%
\frac{1}{\text{Pe}}\frac{\pi (37+21\alpha )}{224\alpha }$, where $\alpha
=\xi /D_{r}$. In the limit $k_{B}T\rightarrow 0$, (i.e., no translational
diffusion in the overdamped limit) we obtain the following expression for
the critical density,
\begin{equation*}
\phi _{c}=-\frac{4\left( -64\alpha +\frac{6\pi (\alpha -2)}{\text{Pe}}+\sqrt{%
4096\alpha ^{2}+\frac{81\pi ^{2}\alpha ^{2}}{\text{Pe}^{2}}-\frac{243\pi
^{2}\alpha }{\text{Pe}^{2}}-\frac{1440\pi \alpha ^{2}}{\text{Pe}}+\frac{%
864\pi \alpha }{\text{Pe}}}\right) }{224\alpha +\frac{\pi (48-15\alpha )}{%
\text{Pe}}},
\end{equation*}%
which again for high enough Peclet number goes as $\phi _{c}\sim \text{Pe}%
^{-1}$. This $1/\text{Pe}$ behavior has also been seen through
phenomenological theories \cite{Cates2014a, marchetti2016} and in kinetic
estimates of the critical density for the onset of MIPS  \cite{Redner2013}.
Finally in fig.\ref{fig.4}, we compare the estimate provided by this theory against
the data from simulations of a system of ABPs interacting through a WCA
potential treating $\alpha =\frac{\xi }{D_{r}}$ as a fit parameter and we
find good agreement with the data, again illustrating the validity of the
result to systems with short range strongly repulsive interactions.

\section{Summary \& Discussion}

In this paper we have provided a derivation of the statistical mechanics for
self-propelled hard disks starting from first principles. We then considered
two applications of the derived statistical mechanics. First is computing
the steady state mechanical pressure of a hard sphere active fluid. In the
athermal limit ($k_{B}T=0$) we find good agreement with existing numerical
simulations \cite{Solon2014} of ABPs for low to moderate densities. In the
absence of activity ($v_{0}=0$) we reproduce the pressure one finds for
thermal hard spheres by using Revised Enskog theory \cite{ap1}. We then
derived the hydrodynamic equations describing self-propelled hard spheres
and identified the critical density for the onset of MIPS in this system. We
find that our prediction  fits well the numerical data \cite{Redner2013}
from ABP simulations and agrees with earlier phenomenological estimates \cite%
{Redner2013, marchetti2016, Cates2014a} in the large Peclet number regime.

This work uses the theory of hard spheres to circumvent the finite collision
time problem and hence many-body effects present in arbitrary repulsive
potentials. While an idealization, the results obtained in this paper should
be useful for any system of self-propelled particles interacting through
strongly repulsive potentials and can be useful for incorporating excluded
volume effects into diverse models for active systems. This can be seen in
the two applications presented, where we have obtained good agreement with
numerical simulations of the overdamped dynamics of ABPs interacting the WCA
potential. The statistical mechanics presented here transcends to the two
applications we have used to illustrate its utility and is potentially
useful in diverse active materials modeling.

\section*{Acknowledgments}

We thank Yaouen Fily and Gabe Redner for providing the numerical data. We acknowledge support from the NSF DMR-1149266, the Brandeis MRSEC DMR-1420382, and  IGERT DGE-1068620.

\appendix

\section{Derivation of the Smoluchowski equation}

We provide a complete derivation of the overdamped dynamics of
self-propelled hard disks. Our starting point will be the coupled Langevin
equations
\begin{equation}
m\partial _{t}\mathbf{v}_{i}=m\sum_{j\neq i}T(i,j)\mathbf{v}_i-\xi \mathbf{v}_{i}+F\hat{%
\mathbf{u}}_{i}+\boldsymbol{\eta }_{i}
\end{equation}
\begin{equation}
\partial _{t}\omega _{i}=-\xi _{R}\omega _{i}+\eta _{i}^{R}
\end{equation}%
where $T(i,j)$ is the binary collision operator given in Eq.(\ref{3}). The noise
function $\boldsymbol{\eta }_{i}$ and $\eta _{i}^{R}$ are both Gaussian
white noise variables with zero mean and the same correlations defined in the main body of text . The phase space variables associated with the above Langevin
dynamics are the position $\mathbf{r}_{n}(t)$, the velocity $\mathbf{v}%
_{n}(t)$, the angular velocity $\boldsymbol{\omega }_{n}(t)$ and the
orientation $\theta _{n}(t)$ of the n particles. We begin by considering an
arbitrary phase space function $A(\Gamma )$ evaluated at some phase point $%
\Gamma _{t}=\left\{ \mathbf{r}_{1}(t),\dots ,\mathbf{r}_{n}(t),\mathbf{v}%
_{1}(t),\dots ,\mathbf{v}_{n}(t),\boldsymbol{\omega }_{1}(t),\dots ,%
\boldsymbol{\omega }_{n}(t),\theta _{1}(t),\dots ,\theta _{n}(t)\right\} $
which has evolved from an initial configuration $\Gamma =\left\{ \mathbf{r}%
_{1},\dots ,\mathbf{r}_{n},\mathbf{v}_{1},\dots ,\mathbf{v}_{n},\boldsymbol{%
\omega }_{1},\dots ,\boldsymbol{\omega }_{n},\theta _{1},\dots ,\theta
_{n}\right\} $. A phase space function at some time t can be expressed in terms
of a generator for the dynamics in the following way
\begin{equation}
A(\Gamma _{t})=e^{Lt}A(\Gamma ).
\end{equation}%
In the above equation, $L$ is the sum of generators for translations and rotations for each
particle coordinate. For a system with pairwise additive, conservative,
non-singular interactions the operator L can be written as
\begin{equation}
L=L_{0}+\frac{1}{2}\sum_{i,j\neq i}m^{-1}\mathbf{F}(r_{ij})\cdot (\nabla _{%
\mathbf{v}_{i}}-\nabla _{\mathbf{v}_{j}}),
\end{equation}%
where the single particle component for self propelled particles with a
propulsion force $F^{sp}$ along its body axis is given by
\begin{equation}
\begin{split}
L_{0}=& \sum_{i}^{N}\{\mathbf{v}_{i}\cdot \nabla _{\mathbf{r}_{i}}+\omega
_{i}\partial _{\theta _{i}}+\frac{F^{sp}}{m}\hat{\mathbf{u}}\cdot \nabla _{%
\mathbf{v}_{i}}-\frac{\xi }{m}\mathbf{v}_{i}\cdot \nabla _{\mathbf{v}%
_{i}}-\xi _{R}\omega _{i}\partial _{\omega _{i}} \\
& +\frac{1}{m}\boldsymbol{\eta }_{i}\cdot \nabla _{\mathbf{v}_{i}}+\eta
_{i}^{r}\partial _{\omega _{i}}-\frac{k_{B}T\xi }{m^{2}}\nabla _{\mathbf{v}%
_{i}}^{2}-\frac{k_{B}T_{R}\xi _{R}}{I}\partial _{\omega _{i}}^{2}\}. \label{A5}
\end{split}%
\end{equation}%
The inclusion of the last two terms in the above generator results
from using Ito calculus to correctly incorporate the stochastic component of
the dynamics. Eq.(\ref{A5}) is the generalization of the completely
deterministic case, in which the dynamics are governed by Hamilton's
equations. One can verify that, for Hamiltonian systems, the generator L
changes the positions and velocities according to Hamilton's equations. In
this case, L is the linear operator representing the Poisson bracket of its
operand with the Hamiltonian for the system \cite{ap2}. However, we are
considering the stochastic dynamics of elastic hard spheres, thus the
interactions must be taking into account via the binary collision operator $%
T(i,j)$. In this case the operator L becomes
\begin{equation}
L=L_{0}+\frac{1}{2}\sum_{i,j\neq i}T(i,j).
\end{equation}%
The above formalism completely determines the stochastic dynamics of any
observable $A(\Gamma )$ for given initial conditions in phase space. In this study, we
would like to write down not the stochastic observable, but its ensemble
averaged value for a given ensemble of initial conditions $\hat{\rho}(\Gamma
)$. This is represented by the following phase space average
\begin{equation}
\langle A(\Gamma )\rangle _{ens}=\int d\Gamma \hat{\rho}(\Gamma )A(\Gamma ,t).
\end{equation}%
One can equivalently treat the phase space density as the dynamical variable 
\cite{zwanzig}
\begin{equation}
\langle A(\Gamma )\rangle _{ens}=\int d\Gamma \hat{\rho}(\Gamma ,t)A(\Gamma ),
\end{equation}%
and taking the time derivative of both equation yields
\begin{equation*}
\int d\Gamma \partial _{t}\hat{\rho}(\Gamma ,t)A(\Gamma )=\int d\Gamma \hat{%
\rho}(\Gamma )\partial _{t}A(\Gamma ,t)
\end{equation*}%
\begin{equation*}
=\int d\Gamma \hat{\rho}(\Gamma )LA(\Gamma ,t)
\end{equation*}%
\begin{equation*}
=-\int d\Gamma \mathcal{L}\hat{\rho}(\Gamma )A(\Gamma ,t).
\end{equation*}%
Where $\mathcal{L}$ is the adjoint operator to L. The result of this simple manipulation is a
Liouville-like equation for the phase space probability density
\begin{equation}
(\partial _{t}+\mathcal{L})\hat{\rho}(\Gamma ,t)=0,
\end{equation}%
with the adjoint operator is given by
\begin{equation}
\begin{split}
\mathcal{L}=& \sum_{i}^{N}\{\mathbf{v}_{i}\cdot \nabla _{\mathbf{r}%
_{i}}+\omega _{i}\partial _{\theta _{i}}+\frac{F^{sp}}{m}\hat{\mathbf{u}}%
\cdot \nabla _{\mathbf{v}_{i}}-\frac{\xi }{m}\nabla _{\mathbf{v}_{i}}\cdot
\mathbf{v}_{i}-\xi _{R}\partial _{\omega _{i}}\omega _{i} \\
& +\frac{1}{m}\boldsymbol{\eta }_{i}\cdot \nabla _{\mathbf{v}_{i}}+\eta
_{i}^{r}\partial _{\omega _{i}}-\frac{k_{B}T\xi }{m^{2}}\nabla _{\mathbf{v}%
_{i}}^{2}-\frac{k_{B}T_{R}\xi _{R}}{I}\partial _{\omega _{i}}^{2}\}-\frac{1}{%
2}\sum_{i,j\neq i}\bar{T}(i,j). \label{A10}
\end{split}%
\end{equation}%
In Eq.(\ref{A10}), the single particle component has been identified by an integration by
parts, while the collision operator $\bar{T}(i,j)$ has been explicitly
constructed using the restituting collisions \cite{aparna2010}
\begin{equation}
\bar{T}(i,j)=\sigma \int d\hat{\boldsymbol{\sigma }}\Theta (\mathbf{V}%
_{ij}\cdot \hat{\boldsymbol{\sigma }})(\mathbf{V}_{ij}\cdot \hat{\boldsymbol{%
\sigma }})[b_{ij}^{-1}\delta (\mathbf{r}_{ij}-\boldsymbol{\sigma }%
)-\delta (\mathbf{r}_{ij}+\boldsymbol{\sigma })].
\end{equation}%
In the above, $b_{ij}^{-1}$ is the generator of restituting collisions $%
(b_{ij}^{-1}A(x_{i}^{\prime },x_{j}^{\prime })=A(x_{i},x_{j}))$ which replaces
post collisional velocities with its pre-collisional values. Finally, we
average over the noise $\rho =\langle \hat{\rho}\rangle$. The resulting equation is given by
\begin{equation}
(\partial _{t}+\mathcal{L})\rho (\Gamma ,t)=0,
\end{equation}%
with the operator $\mathcal{L}$ given by
\begin{equation}
\begin{split}
\mathcal{L}=& \sum_{i}^{N}\{\mathbf{v}_{i}\cdot \nabla _{\mathbf{r}%
_{i}}+\omega _{i}\partial _{\theta _{i}}+\frac{F^{sp}}{m}\hat{\mathbf{u}}%
\cdot \nabla _{\mathbf{v}_{i}}-\frac{\xi }{m}\nabla _{\mathbf{v}_{i}}\cdot
\mathbf{v}_{i}-\xi _{R}\partial _{\omega _{i}}\omega _{i} \\
& -\frac{k_{B}T\xi }{m^{2}}\nabla _{\mathbf{v}_{i}}^{2}-\frac{k_{B}T_{R}\xi
_{R}}{I}\partial _{\omega _{i}}^{2}\}-\frac{1}{2}\sum_{i,j\neq i}\bar{T}(i,j).
\end{split}%
\end{equation}%
The above equation governs the time evolution of the phase space density of
N self-propelled hard spheres. To proceed, we now introduce reduced
distribution functions
\begin{equation}
f^{(n)}(\Gamma _{1},\dots ,\Gamma _{n},t)=\frac{N!}{(N-n)!}\int d\Gamma
_{n+1}\dots d\Gamma _{N}\rho (\Gamma _{1}\dots ,\Gamma _{N},t)
\end{equation}%
and consider the first equation of the resulting hierarchy
\begin{equation}
\partial _{t}f^{(1)}(\Gamma _{1},t)+\mathcal{D}f^{(1)}(\Gamma _{1},t)=\int
d\Gamma _{2}\bar{T}(1,2)f^{(2)}(\Gamma _{1},\Gamma _{2},t). \label{A15}
\end{equation}%
Where the one particle operator $\mathcal{D}$ in Eq.(\ref{A15}) is given by
\begin{equation}
\mathcal{D}=\mathbf{v}_{i}\cdot \nabla _{\mathbf{r}_{i}}+\boldsymbol{\omega }%
_{i}\cdot \partial_{\theta_i}+\frac{F^{sp}}{m}\hat{\mathbf{u}}\cdot \nabla _{%
\mathbf{v}_{i}}-\frac{\xi }{m}\nabla _{\mathbf{v}_{i}}\cdot \mathbf{v}%
_{i}-\xi _{R}\partial _{\omega _{i}}\omega _{i}-\frac{k_{B}T\xi }{m^{2}}%
\nabla _{\mathbf{v}_{i}}^{2}-\frac{k_{B}T_{R}\xi _{R}}{I}\partial _{\omega
_{i}}^{2}.
\end{equation}%
The goal of this section is to obtain a dynamical equation for the local concentration field
by
\begin{equation}
c^{(1)}(\mathbf{r}_{1},\theta _{1},t)=\int d\mathbf{v}_{1}f^{(1)}(\mathbf{r}%
_{1},\mathbf{v}_{1},\theta _{1},\boldsymbol{\omega }_{1},t).
\end{equation}%
For convenience, we introduce a translational and rotational current defined as the
velocity moments of the 1-particle distribution function
\begin{equation}
\mathbf{J}^{T}=\int d\omega _{1}d\mathbf{v}_{1}\mathbf{v}_{1}f^{(1)}(\mathbf{%
r}_{1},\mathbf{v}_{1},\theta _{1},\boldsymbol{\omega }_{1},t)
\end{equation}%
\begin{equation}
J^{R}=\int d\omega _{1}d\mathbf{v}_{1}\omega _{1}f^{(1)}(\mathbf{r}_{1},%
\mathbf{v}_{1},\theta _{1},\boldsymbol{\omega }_{1},t).
\end{equation}%
Taking velocity moments of Eq.(\ref{A15}),  we arrive at the following equations
\begin{equation}
\partial _{t}c^{(1)}+\nabla _{\mathbf{r}_{1}}\cdot \mathbf{J}^{T}+\partial
_{\theta _{1}}J^{R}=0 \label{A20}
\end{equation}%
\begin{equation}
\partial _{t}J_{\alpha }^{T}+\frac{\xi }{m}J_{\alpha }^{T}-\frac{F^{sp}}{m}%
\hat{u}_{1\alpha }c^{(1)}+\partial _{\mathbf{r}_{1}\beta }\langle v_{1\alpha
}v_{1\beta }\rangle +\partial _{\theta _{1}}\langle \omega _{1}v_{1\alpha
}\rangle =I_{\alpha }^{T} \label{A21}
\end{equation}%
\begin{equation}
\partial _{t}J^{R}+\xi _{R}J^{R}+\partial _{\mathbf{r}_{1}\beta }\langle
\omega _{1}v_{1\beta }\rangle +\partial _{\theta _{1}}\langle \omega
_{1}^{2}\rangle =I^{R}. \label{A22}
\end{equation}%
In the above,
\begin{equation}
\langle v_{1\alpha }v_{1\beta }\rangle =\int d\omega _{1}d\mathbf{v}%
_{1}v_{1\alpha }v_{1\beta }f^{(1)}(\mathbf{r}_{1},\mathbf{v}_{1},\theta _{1},%
\boldsymbol{\omega }_{1},t)
\end{equation}%
\begin{equation}
\langle \omega _{1}v_{1\alpha }\rangle =\int d\omega _{1}d\mathbf{v}%
_{1}\omega _{1}v_{1\alpha }f^{(1)}(\mathbf{r}_{1},\mathbf{v}_{1},\theta _{1},%
\boldsymbol{\omega }_{1},t)
\end{equation}%
\begin{equation}
\langle \omega _{1}^{2}\rangle =\int d\omega _{1}d\mathbf{v}_{1}\omega
_{1}^{2}f^{(1)}(\mathbf{r}_{1},\mathbf{v}_{1},\theta _{1},\boldsymbol{\omega
}_{1},t),
\end{equation}%
are the second velocity moments of the probability distribution function. Also present are the following terms
\begin{equation}
I_{\alpha }^{T}=\int d\Gamma _{2}\int d\omega _{1}d\mathbf{v}_{1}v_{1\alpha }%
\bar{T}(1,2)f^{(2)}(\Gamma _{1},\Gamma _{2},t) \label{A26}
\end{equation}%
\begin{equation}
I^{R}=\int d\Gamma _{2}\int d\omega _{1}d\mathbf{v}_{1}\omega _{1}\bar{T}%
(1,2)f^{(2)}(\Gamma _{1},\Gamma _{2},t)=0.\label{A27}
\end{equation}%
Eq.(\ref{A26}) arises from the linear momentum transfer due to collisions. The integral in Eq.(\ref{A27}) equates to zero because these are smooth hard discs and therefore, no angular momentum is transferred in a collision. The
translational and rotational currents in the above equations are subject to
frictional damping and as such relax on time scales of order $m/\xi $. On
time scales $t\gg m/\xi $ the flux can be approximated as
\begin{equation}
\lim_{t\gg m/\xi }J_{\alpha }^{T}=-\frac{m}{\xi }\big[-\frac{F^{sp}}{m}\hat{u%
}_{1\alpha }+\partial _{\mathbf{r}_{1}\beta }\langle v_{1\alpha }v_{1\beta
}\rangle +\partial _{\theta _{1}}\langle \omega _{1}v_{1\alpha }\rangle
-I_{\alpha }^{T}\big] \label{A28}
\end{equation}%
\begin{equation}
\lim_{t\gg 1/\xi _{R}}J_{=}^{R}-\frac{1}{\xi _{R}}\big[\partial _{\mathbf{r}%
_{1}\beta }\langle \omega _{1}v_{1\beta }\rangle +\partial _{\theta
_{1}}\langle \omega _{1}^{2}\rangle \big] \label{A29}
\end{equation}

\subsection*{Velocity Integration}

 To complete the derivation of the Smoluchowski equation, we must evaluate the velocity integrals in Eqs.(\ref{A28}-\ref{A29}). As outlined in section II of the main text, we assume that on these timescales the velocity
distributions have relaxed to their local equilibrium form. Explicitly we have that 
\begin{equation}
f(\mathbf{v})f(\omega)=Ne^{\frac{-m}{2k_BT}(\mathbf{v}-v_0\hat{\mathbf{u}}%
)^2}e^{-\frac{I\omega^2}{2k_BT_R}},
\end{equation}
where N is a normalization factor. With this distribution, it is readily shown that
\begin{equation}
\langle \omega_1v_{1\alpha}\rangle=\langle \omega_1v_{1\beta}\rangle=0
\end{equation}
\begin{equation}
\langle\omega_1^2\rangle=\frac{k_BT}{I}
\end{equation}
\begin{equation}
\langle v_\alpha v_\beta \rangle=(v_0^2+\frac{k_BT}{m})\hat{u}_\alpha\hat{u}%
_\beta+\frac{k_BT}{m}(\delta_{\alpha\beta}-\hat{u}_\alpha\hat{u}_\beta).
\end{equation}
The above averages are exact for the one-body terms, but for the collision
integral (which depends on the orientation of the colliding particles and an
integration over the two body distribution), the exact evaluation of the
velocity integrals is not possible. To continue further, we make the following
asymptotic approximation that accurately captures the physics in the limits
that $v_0^2\gg k_BT/m$ (or $k_BT/m\gg v_0^2$)
\begin{equation}
f(v)=\frac{1}{2\pi k_BT m^{-1}+v_0^2}(e^{-\frac{mv^2}{2k_BT}}+v_0^2\delta(%
\mathbf{v}-v_0\hat{\mathbf{u}})).
\end{equation}
The delta function enforces that the particle will have a self-propulsion
velocity $v_0$ along its body axis $\hat{\mathbf{u}}$ and reproduces the the
correct distribution in the athermal limit. The Maxwellian accounts for
thermal noise and reproduces the statistical mechanics of thermal hard
spheres in the limit of zero self-propulsion.

The last step is to perform the velocity averaging in Eq.(\ref{A26}). Explicitly we
must evaluate the following,
\begin{equation}
\int d\mathbf{v}_1d\mathbf{v}_2v_{1\alpha}\bar{T}(1,2)f(\mathbf{v}_1)f(%
\mathbf{v}_2).
\end{equation}
This evaluation of this integral leads to Eqs.(\ref{12a}-\ref{12c}) for $%
I^{thermal}_\alpha $, $I^{active}_\alpha$, and $I^{cross}_\alpha$. To
evaluate $I^{cross}_\alpha $, which comes from the cross terms in the
multiplication of the velocity distributions, one must make an additional
approximation. Contained within the integral for the cross terms, one finds
terms like
\begin{equation}
\int d\mathbf{v}_1\Theta(\mathbf{v}_1\cdot\hat{\boldsymbol{\sigma}}-v_0\hat{%
\mathbf{u}}_2\cdot\hat{\boldsymbol{\sigma}}).
\end{equation}
Since the integral is over all possible velocities, to good approximation we
have
\begin{equation}
\int d\mathbf{v}_1\Theta(\mathbf{v}_1\cdot\hat{\boldsymbol{\sigma}}-v_0\hat{%
\mathbf{u}}_2\cdot\hat{\boldsymbol{\sigma}})\approx \int d\mathbf{v}_1\Theta(%
\mathbf{v}_1\cdot\hat{\boldsymbol{\sigma}}).
\end{equation}
Using this approximation, the integrals can be computed. The integrals leading
to $I^{thermal}_\alpha$ and $I^{active}_\alpha$ can be computed exactly.
These are the most important contributions since, in the limits of purely
self-propelled or purely thermal, these are the only contributing
factors. This completes the derivation of the Smoluchowski equation.
Combining Eq.(\ref{A20}), Eqs.(\ref{A28}-\ref{A29}) and the velocity averages yields Eqs.(\ref{10}-\ref{11}) in the main text.

\section{Derivation of Pressure}

To evaluate this expression we start with the Smoluchowski equation (Eq.(\ref{10})).
In steady state, the Smoluchowksi equation takes the following form
\begin{equation}
\nabla \cdot \big[v_0\hat{\mathbf{u}}c-\xi^{-1}\nabla V(\mathbf{r}) c-%
\tensor{\mathbf{D}}\cdot \nabla c+ \xi^{-1}\big(\mathbf{I}^{thermal} +%
\mathbf{I}^{sp}+\mathbf{I}^{cross}\big) \big]+D_r\partial_\theta^2 c=0.
\label{B1}
\end{equation}
Because of the translational invariance in $\hat{y}$, in steady state, the
only spatial dependence in the above is through the x-coordinate.
Integrating Eq.(\ref{B1}) over the orientations $\theta$, on can see clearly that
the resulting equation is of the form $\partial_x J^T=0$ with $J^T$ being
the particle current. Since this system has impermeable boundary
conditions ($J^T=0$ at the wall), the only admissible solution is that $%
J^T=0 $ everywhere. Written explicitly we have,
\begin{equation}
\int d\theta c \partial_x V= \xi \int d \theta \big[v_{0}\hat{u}_x c-%
D_{xx}\partial_x c+\xi^{-1}(I^{thermal}_x +I^{sp}_x+{I}^{cross}_x)%
\big].  \label{B2}
\end{equation}
Rewriting the left hand side of Eq.(\ref{B2}) in terms of the density by noting that
$\rho=\int c d\theta$ and by integrating over x, we obtain the pressure (Eq.(\ref{14})) written in terms of the concentration field
\begin{equation}
P=\int_0^\infty dx \int d\theta \xi[v_0 \hat{u}_{x}c-D%
_{xx}\partial_x c+\xi^{-1}(I^{thermal}_x+I^{sp}_x+I^{cross}_x)].  \label{B3}
\end{equation}
Now let us multiply Eq.(\ref{B1}) by $\hat{u}_x$ and integrate over $\theta$
\begin{equation}
D_r \int d\theta \hat{u}_xc=-\partial_x \int d\theta \big[\hat{u}_x\hat{u}_x
v_0c-\xi^{-1}\hat{u}_x c\partial_x V-\hat{u}_x D_{xx} \partial_x
c+\xi^{-1}\hat{u}_{x}(I^{thermal}_x +I^{sp}_x+I^{cross}_x)\big],  \label{B4}
\end{equation}
and now integrate Eq.(\ref{B4}) over x. Note that the right hand side of Eq.(\ref{B4}) is given as a total derivative and therefore trivially integrated. Aside from the interaction terms in Eq.(\ref{B4}), the only surviving term is the angular integral over $\hat{u}_x\hat{u}_x c$ which is proportional $\rho/2$. The remaining terms vanish because there can be no orientational order in the bulk of the fluid $(x=0)$ or at $x=\infty$. With these facts, we have that
\begin{equation}
D_r \int_0^\infty dx \int d\theta \hat{u}_xc= \frac{v_0}{2}\rho_0+\int
d\theta\xi^{-1}\hat{u}_{x}(I^{thermal}_x +I^{sp}_x+I^{cross}_x)\bigg|_{x=0}.
\label{B5}
\end{equation}
Using Eq.(\ref{B5}) to eliminate the first term on the right hand side of Eq.(\ref{B3})
we recover Eq.(\ref{P1})
\begin{equation}
P=P_0+\int d\theta \bigg[\frac{v_0}{D_r} \hat{u}%
_x(I^{thermal}_x+I^{sp}_x+I^{cross}_x)\bigg|_{x=0}+\int_0^\infty
(I^{thermal}_x+I^{sp}_x+I^{cross}_x)dx\bigg].
\end{equation}

\section{Evaluation of Mean-Field Force and Low Moments Closure}
In this section we give the explicit evaluation of the mean-field force (Eqs.(\ref{12a}-\ref{12c}))
and an outline of the low moment closure procedure used in the text. To
construct the low moment closure we first represent the concentration field
as the following harmonic expansion \cite{aphrodite}
\begin{equation}
c(\mathbf{r},\theta,t)=\sum_{m=0}^\infty a^m_{i_1\dots i_m}(\mathbf{r}%
,t)T^m_{i_1\dots i_m}(\theta),
\end{equation}
where the irreducible tensors $T^m_{i_1\dots i_m}(\theta)$ are equivalent to
the spherical harmonics but expressed here in Cartesian coordinates. For illustrative purposes, the
first four irreducible tensors are given by
\begin{equation}
T^0=1
\end{equation}
\begin{equation}
T^1_i=\hat{u}_i
\end{equation}
\begin{equation}
T^2_{ij}=\hat{u}_i\hat{u}_j-\frac{1}{2}\delta_{ij}
\end{equation}
\begin{equation}
T^3_{ijk}=\hat{u}_i\hat{u}_j\hat{u}_k-\frac{1}{4}(\delta_{ij}\hat{u}%
_k+\delta_{ik}\hat{u}_j+\delta_{jk}\hat{u}_i).
\end{equation}
The $m$ th order moment $a^m_{i_1\dots i_m}(\mathbf{r},t)$ of the concentration is given by
\begin{equation}
a^m_{i_1\dots i_m}(\mathbf{r},t)=\frac{1}{2\pi}\int d\theta T^m_{i_1\dots
i_m}(\theta)c(\mathbf{r},\theta,t).
\end{equation}
The dynamical equation for the $m$ th moment can then be obtained by taking moments of the Smoluchowksi equation (Eq.(\ref{10})), with the result
\begin{equation}
\partial_ta^m_{i_1\dots i_m}(\mathbf{r},t)=-m^2D_ra^m_{i_1\dots i_m}(\mathbf{r},t)-\partial_\alpha\int d\theta T^m_{i_1\dots i_m}(\theta)J^T_\alpha. \label{moments}
\end{equation}
We see that when written in this form, the density and polarization are just given as the
zeroth and first moment of the concentration field. The low moment closure
is the approximation that $c(\mathbf{r}_1,\theta_1,t)$ can be expressed as a
functional of the first two moments $c^{(1)}(\theta_1;\{\rho(\mathbf{r}%
,t),P_\alpha(\mathbf{r},t)\})$. The motivation for this closure is the following: One can see from Eq.(\ref{moments}) that all moments greater than the zeroth moment have a finite relaxation rate given by $m^2 Dr$. Thus, these moments will relax to values that depend on gradients of the local concentration. When the values of the higher moments are substituted into the density equation they result in terms that are irrelevant compared to the terms resulting from the polarization equation, i.e., contain more powers of gradients and fields. Therefore, we close the expansion by
assuming that the second and higher moments can be neglected. This implies
that $\int d\theta\hat{u}_i\hat{u}_jc=\frac{1}{2}\delta_{ij}\rho$ and $\int
d\theta \hat{u}_i\hat{u}_j\hat{u}_kc=\frac{1}{4}(\delta_{ij}P_k+%
\delta_{ik}P_j+\delta_{jk}P_i)$. We note that this is a valid closure in the absence of aligning interactions because the density and polarization represent the relevant macroscopic fields. To accurately
capture the effects of aligning interactions one must include the second
moment as is done in \cite{farrell, aparnarod, ben1}. With this we can now
evaluate the mean field force. The mean field force consists of three parts $%
I_\alpha^{thermal}$, $I_\alpha^{sp}$, and $I_\alpha^{cross}$. To evaluate
these quantities we first make the functional ansatz used in the main body
of the text, namely that $c^{(2)}(\mathbf{r}_1,\theta_1,\mathbf{r}_1-%
\boldsymbol{\sigma},\theta_2,t)=g(\sigma)c(\mathbf{r}_1,\theta_1,t)c(\mathbf{%
r}_1-\boldsymbol{\sigma},\theta_2,t)$. We then gradient expand the nonlocal
dependence of the concentration field $c(\mathbf{r}-\boldsymbol{\sigma}%
,\theta_2,t)=c(\mathbf{r},\theta_2,t)-\sigma_\alpha\partial_\alpha c(\mathbf{%
r},\theta_2,t)+\dots$ , which is valid in a long wavelength (hydrodynamic)
description of the system. To first order in gradients the integrals can be
evaluated. The result of this integration is 
\begin{equation}
I_\alpha^{thermal}=-\frac{\sigma^2 4\pi^2(k_BT)^3}{(2\pi k_BT+v_0^2)^2}%
g(\sigma)\partial_\alpha\rho(\mathbf{r},t)c(\mathbf{r}_1,\theta_1,t)
\end{equation}
\begin{equation}
I_\alpha^{sp}=g(\sigma)\frac{\sigma v_0^6}{(2\pi k_BT+v_0^2)^2}\bigg[-\frac{4%
}{3}\hat{u}_{1\alpha}\rho+\frac{4}{3}P_\alpha-\frac{\pi \sigma}{4}\big(\hat{u%
}_{1\alpha}\hat{u}_{1\beta}\partial_\beta \rho+\hat{u}_{1\beta}\partial_%
\beta P_\alpha+\hat{u}_{1\alpha}\partial_\beta P_\beta-\partial_\alpha \rho %
\big) \bigg]c(\mathbf{r}_1,\theta_1,t)
\end{equation}
\begin{equation}
I_\alpha^{cross}=g(\sigma)\frac{\sigma k_BT}{(2\pi k_BT+v_0^2)^2}\bigg[%
2\pi(k_BT)^{1/2}v_0^3\big(P_\alpha-\hat{u}_{1\alpha}\rho\big)-\sigma\big(%
2\pi^2k_BTv_0^2+\frac{8+\pi}{4}\pi v_0^4\big)\partial_\alpha \rho
\end{equation}
\begin{equation*}
+\frac{\sigma\pi^2}{2}v_0^4\hat{u}_{1\alpha}\hat{u}_{1\beta}\partial_\beta%
\rho\bigg]c(\mathbf{r}_1,\theta_1,t).
\end{equation*}

\section{Constants and functions used in the hydrodynamics and pressure}

For convenience, let us call $A=(2\pi k_{B}T+v_{0}^{2})^{-1}$ and $a=(1+%
\frac{v_{0}^{2}}{2k_{B}T})$. The constants in the hydrodynamic equations and pressure are
then given by
\begin{equation}
D_{p}=\frac{1}{\xi }\frac{\pi }{4}g(\sigma )\sigma ^{2}A^{2}v_{0}^{6}
\end{equation}%
\begin{equation}
\lambda _{I}=\frac{1}{2\xi }A^{2}\sigma g(\sigma )[\frac{4}{3}%
v_{0}^{6}+v_{0}^{3}(2\pi k_{B}T)^{3/2}]
\end{equation}%
\begin{equation}
\lambda _{D}=\xi (\mathcal{D}(\rho _{0})-\frac{D_{\parallel }}{2}-D_{\perp }\big)
\end{equation}%
\begin{equation}
\lambda _{2}=\frac{D_{p}}{4}+\frac{1}{3}v_{0}^{4}\sigma ^{2}A^{2}\pi \frac{%
k_{B}T}{\xi }g(\sigma )
\end{equation}%
and the functions are given by
\begin{equation}
K(\rho )=\frac{1}{2}D_{p}\rho
\end{equation}%
\begin{equation}
L_{\alpha \beta }(P_{\gamma })=\frac{A^{2}k_{B}Tg(\sigma )\sigma ^{2}}{\xi }%
P_{\gamma }\delta _{\alpha \beta }\bigg[a(2\pi k_{B}T)^{2}+(\frac{3\pi ^{2}}{%
8}+2)v_{0}^{4}+\frac{(2+3\pi )\pi -16}{4\pi }\frac{v_{0}^{6}\pi }{k_{B}T}%
\bigg]
\end{equation}%
\begin{equation}
\mathcal{D}(\rho )=\big(\frac{D_{\parallel }}{2}+D_{\perp }\big)+\bigg(a(2\pi
k_{B}T)^{2}+\frac{3}{8}\frac{v_{0}^{6}\pi }{k_{B}T}+\frac{\pi +4}{2}v_{0}^{4}%
\bigg)\frac{A^{2}k_{B}Tg(\sigma )\sigma ^{2}}{\xi }\rho
\end{equation}%

\bibliography{paperbib2}

\end{document}